\documentclass[twocolumn,showpacs,preprintnumbers,amsmath,amssymb]{revtex4}
\usepackage{graphicx}% Include figure files
\usepackage{dcolumn}% Align table columns on decimal point
\usepackage{bm}% bold math

\begin{document}
\preprint{final}

\title{Neutron Diffraction Study of Field Cooling Effects on Relaxor Ferroelectrics Pb[(Zn$_{1/3}$Nb$_{2/3}$)$_{0.92}$Ti$_{0.08}$]O$_{3}$}% Force line breaks with \\

\author{Kenji Ohwada}
 \email{ohwada@spring8.or.jp}
\affiliation{%
Synchrotron Radiation Research Center (at SPring-8), Japan Atomic Energy Research Institute, 679-5148, Japan
}%

\author{Kazuma Hirota}
\affiliation{%
Department of Physics, Tohoku University, Sendai 980-8578, Japan
}%

\author{Paul W.\ Rehrig}
 \altaffiliation[Present address: ]{TRS Ceramics, Inc., 2820 East College Ave. State College, PA 16801, U.S.A. \\}
\affiliation{%
Materials Research Laboratory, The Pennsylvania State University, PA 16802, U.S.A.
}%

\author{Yasuhiko Fujii}
\affiliation{%
Institute for Solid State Physics, The University of Tokyo, Kashiwa 277-8581, Japan
}%

\author{Gen Shirane}
\affiliation{%
Department of Physics, Brookhaven National Laboratory, Upton, NY 11973-5000, U.S.A.
}%

\date{\today}% It is always \today, today,

\begin{abstract}
High-temperature ($T$) and high-electric-field ($E$) effects on Pb[(Zn$_{1/3}$Nb$_{2/3}$)$_{0.92}$Ti$_{0.08}$]O$_{3}$ (PZN-8\%PT) were studied comprehensively by neutron diffraction in the ranges $300 \le T \le 550$~K and $0 \le E \le 15$~kV/cm. We have focused on how phase transitions depend on preceding thermal and electrical sequences. In the field cooling process (FC, $E \parallel [001] \ge 0.5$~kV/cm), a successive cubic (C) $\rightarrow$ tetragonal (T) $\rightarrow$ monoclinic (M$_{\rm C}$) transition was observed. In the zero field cooling process (ZFC), however, we have found that the system does {\em not} transform to the rhombohedral (R) phase as widely believed, but to a new, unidentified phase, which we call X. X gives a Bragg peak profile similar to that expected for R, but the $c$-axis is always slightly shorter than the $a$-axis. As for field effects on the X phase, we found an irreversible X $\rightarrow$ M$_{\rm C}$ transition via another monoclinic phase (M$_{\rm A}$) as expected from a previous report [Noheda {\it et al.}\ Phys.\ Rev.\ Lett. {\bf 86}, 3891 (2001)]. At a higher electric field, we confirmed a $c$-axis jump associated with the field-induced M$_{\rm C}$ $\rightarrow$ T transition, which was observed by strain and x-ray diffraction measurements.
\end{abstract}

\maketitle

%%%
%%% Introduction
%%%

\section{Introduction}

%%%%
%%%% ============================= Fig. 1 ============================== %
%%%%
\begin{figure}[htb]
\begin{center}
\includegraphics[width=0.85\linewidth]{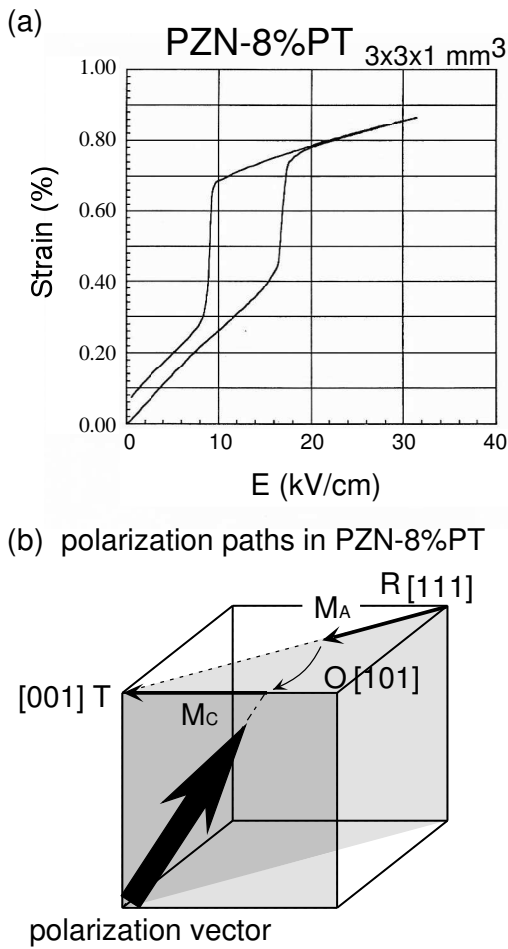}
\end{center}
\caption{(a) Electric field dependence of the strain measured for the 3$\times$3$\times$1 mm$^{3}$ PZN-8\%PT crystal. (b) Polarization rotation path for the PZN-8\%PT crystal (R $\rightarrow$ M$_{\rm{A}}$ $\rightarrow$ M$_{\rm{C}}$ $\rightarrow$ T).}
\label{fig:1}
\end{figure}
%%%% ============================= Fig. 1 ============================== %

  Solid solutions of Pb(Zn$_{1/3}$Nb$_{2/3}$)O$_{3}$ and PbTiO$_{3}$ (PZN-$x$PT) are relaxor ferroelectrics with extremely high piezoelectric responses~\cite{Park1,Liu}, which are an order of magnitude larger than those of conventional piezoelectric ceramics such as Pb(Zr$_{x}$Ti$_{1-x})$O$_{3}$ (PZT)~\cite{Kuwata1,Kuwata2}. PZN-$x$PT has a cubic (C) perovskite-type structure at high temperatures, and undergoes a diffuse ferroelectric phase transition at low temperatures. It has been reported that the ferroelectric region is separated into rhombohedral (R) and tetragonal (T) symmetries by a morphotropic phase boundary (MPB), a nearly vertical line between the two phases, at $x \sim 10$\%. The piezoelectric response of PZN-$x$PT reaches the maximum at $x=8$\%, which is located on the R (lower $x$) side near MPB~\cite{Park1}. These behaviors are qualitatively similar to those of PZT. However, PZN-$x$PT can be grown in single crystal form unlike PZT, so that the structural properties can be studied in more detail.
  
  A typical strain field loop for a poled PZN-8\%PT single crystal at room temperature (RT), where an electric field is applied along the pseudo-cubic $[001]$ direction ($E \parallel [001]$), is shown in Fig.~\ref{fig:1}(a) after Park and Shrout. The strain increases linearly below a certain threshold field, then jumps discontinuously, which is called the $c$-jump. Durbin {\it et al.}\ studied PZN-8\%PT by x-ray diffraction and found that the field dependence of the lattice parameter exactly reflects the strain behavior, which indicates the observed high macroscopic strain levels indeed originate from the microscopic strain of the lattice~\cite{Durbin1}. In addition, they found that the structure of PZN-8\%PT exhibits an irreversible change from the zero-field R phase to different phases by applying electric fields~\cite{Durbin2}. Subsequent x-ray measurements at RT by Noheda {\it et al.} have identified the symmetries of the various phases appearing in PZN-8\%PT~\cite{Noheda1,Noheda2}. They have shown that the zero-field R phase starts to follow the direct polarization path to the T symmetry via an intermediate monoclinic (M$_{\rm A}$) phase~\cite{M}, but then jumps irreversibly to an alternate path involving a different type of monoclinic distortion (M$_{\rm C}$)~\cite{M}, as schematically drawn in Fig.~\ref{fig:1}(b). Their result suggests that there is a very narrow region of monoclinic phase nestled against MPB as found for PZT.
  
  The maximum piezoelectric activity is located in the monoclinic phase near MPB in both PZT and PZN-$x$PT. These observations have resulted in the concept of the polarization rotation mechanism by Fu and Cohen~\cite{Fu}, which successfully explains the ultra-high electromechanical response. Whereas the direction of the polarization vector in a conventional tetragonal ferroelectric phase is fixed to the [001] (e.g., PbTiO$_{3}$) or [111] (e.g., BaTiO$_{3}$) direction, the monoclinic symmetry allows the polarization vector a much greater degree of freedom as it is only constrained to lie within the $(1\bar{1}0)$ plane for M$_{\rm A}$ and the (010) plane for M$_{\rm C}$. In the monoclinic phases, the polarization direction can easily adjust to the electric field, thus naturally gives a large piezoelectric response. Although the presence of a monoclinic phase in a cubic perovskite system seems quite unusual, it is now understood within the framework of an extended Devonshire theory for strongly anharmonic crystals, which require higher order terms~\cite{VC}.
  
  As described above, PZN-8\%PT shows a complicated field-induced phase transitions: R $\rightarrow$ M$_{\rm A}$ $\rightarrow$ M$_{\rm C}$ $\rightarrow$ T on increasing field and T $\rightarrow$ M$_{\rm C}$ on decreasing field~\cite{Noheda1,Noheda2}. To clarify the origin of the exceptional piezoelectric character of this system, it is necessary to resolve the complexities of the transformation sequences in more detail. Ohwada {\it et al.} have carried out neutron diffraction experiments on PZN-8\%PT single crystals as a function of applied electric field~\cite{Ohwada}. They have confirmed the irreversible R $\rightarrow$ M$_{\rm A}$ $\rightarrow$ M$_{\rm C}$ sequence reported in x-ray diffraction experiments. However, the sharp $c$-jump observed in strain and x-ray measurements was not reproduced. Instead, they have found a marked asymmetry of the (002) Bragg peak line shapes along the $c$-axis, indicating a non-uniform strain distribution within the crystal, which washes out the $c$-jump. Their subsequent high-energy x-ray study of the same crystals has supported this view.
  
  In the present study, we have carried out high $q$-resolution neutron diffraction experiments of PZN-8\%PT in the temperature and electric-field ranges $300 \le T \le 550$~K and $0 \le E \le 15$~kV/cm, which are significantly extended than those of the previous work. We have particularly focused upon how phase transitions depend on preceding thermal and electrical sequences. The present article is organized as follows: The experimental details are given in Section~II. In Section~III, we show our experimental results on the structural phase transitions under various sequences. We first clarify that PZN-8\%PT does {\em not} transform to the R phase in the zero-field cooling (ZFC) process as widely believed, but to a yet unidentified phase, which we call the X phase. We then describe the temperature dependence in the field cooling (FC) process, and the electric field dependence at fixed temperatures. These results are summarized in an $E$-$T$ phase diagram. Discussions on the novel X phase and the established $E$-$T$ phase diagram are given in Section~IV.

%%%
%%% Experimental
%%%

\section{Experimental Details }
  Neutron diffraction measurements have been performed mainly on the 8$\times$8$\times$2~mm$^3$ single crystal. This crystal was grown by the flux solution method~\cite{Maureen} at the Pennsylvania State University. The as-grown crystal was poled at 10 kV/cm at RT. The strain curves measured along [001] as a function of electric field were used to check the quality of the crystals. As shown for a typical example in Fig.~\ref{fig:1}, the crystals we studied exhibit a sharp jump in the $c$-axis lattice spacing around 15~kV/cm. 

  The neutron diffraction experiments were carried out on the Tohoku University triple-axis spectrometer TOPAN (6G) installed in the JRR-3M reactor located at the Japan Atomic Energy Research Institute in Tokai, Japan. Most of the experiments were performed using the tight horizontal beam collimation 15$'$-10$'$-S-10$'$-B (S = Sample, B = Blank) with two pyrolytic graphite (PG) filters before and after the sample to eliminate higher harmonics in the neutron beam. The incident neutron energy ($E_{i}$) was tuned to 14.7 meV ($\lambda$ = 2.36 \AA) with a highly oriented PG (HOPG) monochromator.
  Gold electric wires insulated with alumina tubes were carefully connected to the sample which was mounted on the cold head of a high-temperature closed cycle refrigerator (HT-CTI, 15 $\le T \le$ 600 K). A high electric field was generated with the MATSUSADA AR-series high voltage power supply. The sample was wrapped with the glass wool for electric insulation, and fixed softly to a specially designed copper sample holder with care not to give unnecessary stress. 

  The crystal was oriented to give the (H0L) scattering plane and the electrodes were attached to the (001) surfaces. Thus electric field was applied along the pseudo-cubic [001] direction in the present study. The lattice constant of PZN-8\%PT in the cubic phase at $T$ = 540 K, $E$ = 0 kV/cm is $a$ = 4.04 \AA, thus the 1 r.l.u.~\cite{rlu} corresponds to $a^{*}$(= $b^{*}$)= 2$\pi$/a = 1.555 \AA$^{-1}$. We fixed the $a^{*}$ (= $b^{*}$) value at 1.555 \AA$^{-1}$ all the time for reciprocal lattice scanning. Scans ware carried out immediately after changing the sample condition.

%%%
%%% Phase transitions
%%%

%%%%
%%%% ============================= Fig. 2 ============================== %
%%%%
\begin{figure}[t]
\begin{center}
\includegraphics[width=1.00\linewidth]{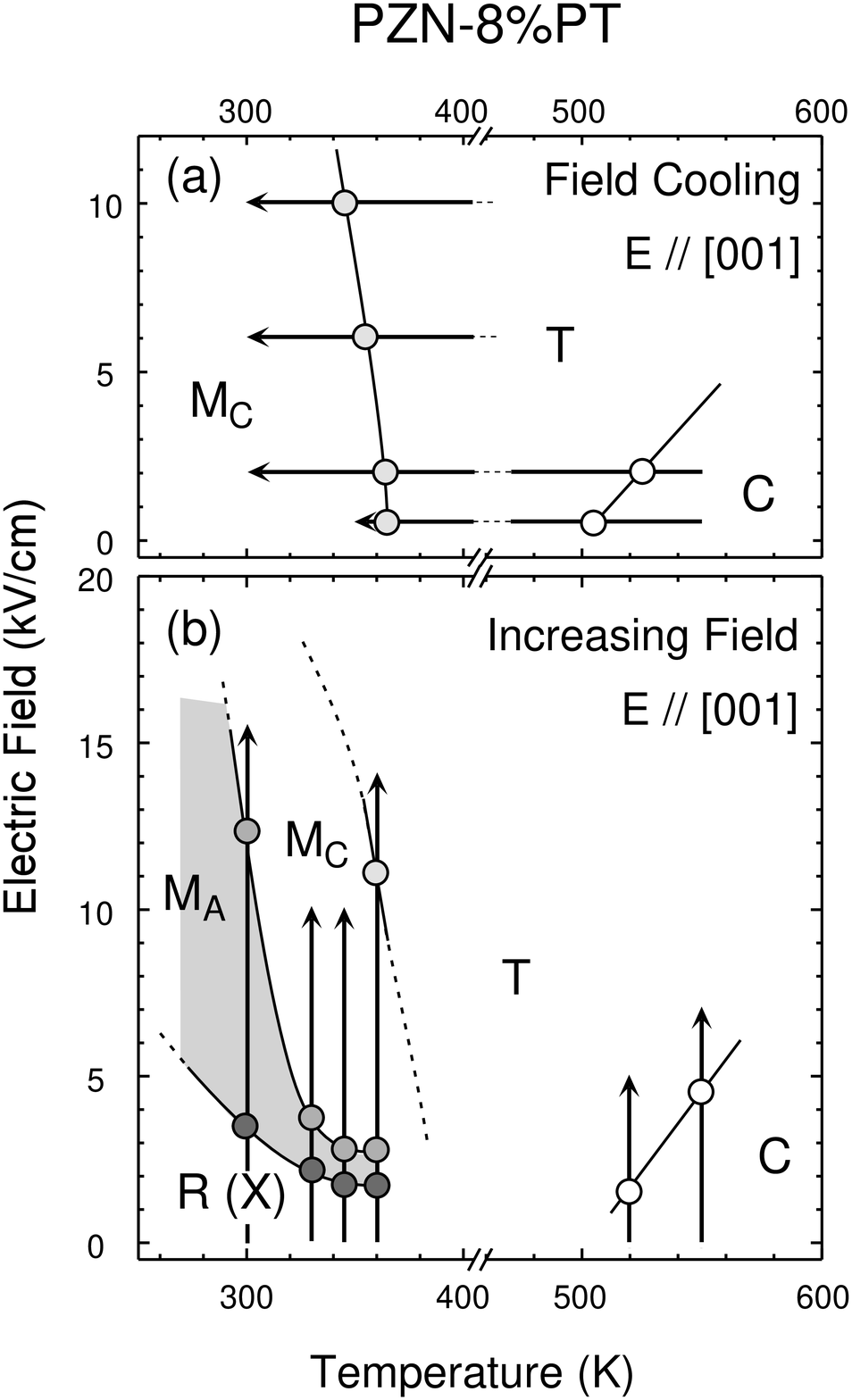}
\end{center}
\caption{
$E$-$T$ phase diagram of PZN-8\%PT. (a) is obtained from structural measurements in the FC process. (b) is obtained from the increasing electric field process after ZFC (and ZFH for the X phase). Arrows indicate the scanning directions and ranges of the corresponding measurement sequences. Circles represent the transition temperatures and fields determined from each sequence.
}
\label{fig:2}
\end{figure}
%%%% ============================= Fig. 2 ============================== %

%%%%
%%%% ============================= Fig. 3 ============================== %
%%%%
\begin{figure}[t]
\begin{center}
\includegraphics[width=0.9\linewidth]{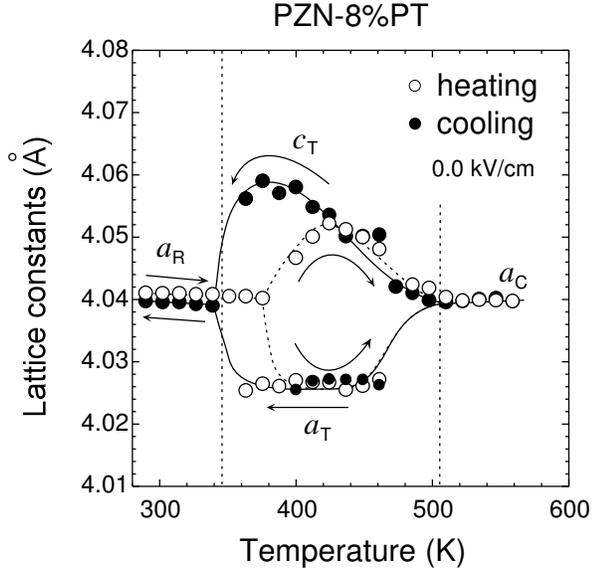}
\end{center}
\caption{
Temperature dependence of the lattice constants under zero-electric field $E$ = 0.0 kV/cm by monitoring the (002) reflection. C $\rightarrow$ T $\rightarrow$ X transition sequence is clearly seen.
}
\label{fig:3}
\end{figure}
%%%% ============================= Fig. 3 ============================== %

%%%%
%%%% ============================= Fig. 4 ============================== %
%%%%
\begin{figure}[t]
\begin{center}
\includegraphics[width=0.9\linewidth]{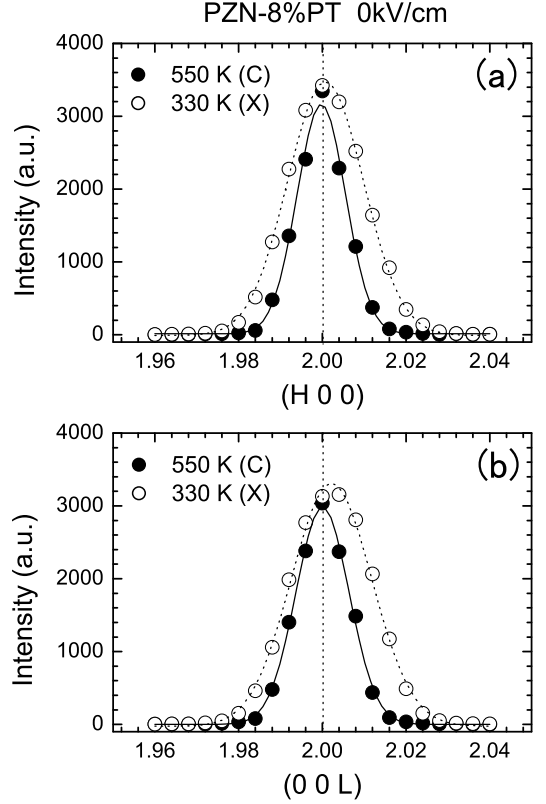}
\end{center}
\caption{
The (H00) and (00L) peak profiles of the C phase and X phase respectively. Solid and dotted lines drawn through the data points are for guides to the eyes.
}
\label{fig:4}
\end{figure}
%%%% ============================= Fig. 4 ============================== %

\section{Phase transitions}
  First of all, we show the Electric field-Temperature ($E$-$T$) phase diagram of PZN-8\%PT in Fig.~\ref{fig:2} which summarizes our present structural measurements.
Circles represent the transition temperatures and fields which were determined by changes of the lattice constants and peak profiles. An arrow indicates the scanning direction and range of the corresponding measurement sequence.

\subsection{X phase}

  We studied the temperature dependence of the lattice constants of PZN-8\%PT crystal under zero electric field (ZF, $E$ = 0.0 kV/cm). The sample was firstly heated up to 560 K, where we confirmed the symmetry is cubic. At 500 K in the ZFC process, the PZN-8\%PT crystal transforms into the T phase from the C phase as expected, associated with 90$^{\circ}$-domain formation, which was confirmed by observing a peak splitting of the (002) reflection along the [001] direction. 
%%%%%%%%%%
%We found that the domain walls are mainly formed along the \{100\} pseudo-cubic planes.
%%%%%%%%%%
  By fitting the (002) reflection with a double Gaussian function, we obtained the temperature dependence of the lattice constants $c_{\rm{T}}$ and $a_{\rm{T}}$ as shown in Fig.~\ref{fig:3}. 

  To confirm the T to R phase transition, we cooled down the sample further. Quite unexpectedly, however, we have found that the system does not transformed into the R phase at 340 K on cooling as widely believed so far, but to an unidentified phase. We will call it the X phase in the present article. Figure~\ref{fig:4} shows the (200) and (002) peak profiles in the C and X phases. The (200) and (002) peak profiles are perfectly centered when $a^{*}$ and $b^{*}$ were set to the same value (1.555 \AA$^{-1}$ in the C phase). On the other hand, the peak center of the (002) profiles in the X phase shifts to higher $q$ compared with that of (200), which indicates PZN-8\%PT is not in the R symmetry after ZFC. 
  We also studied (101) peak profile and found no sign of the expected peak split of the rhombohedral structure, although a mosaic broadening was observed.
  As for the lattice constants, $c_{\rm{T}}$ gradually increases with decreasing temperature and suddenly drops with a large amount at 340 K, where the T $\rightarrow$ X phase transition takes place. 

  Although it is not shown in Fig.~\ref{fig:3}, we also observed the intensity jump at the transition temperatures, 500 K and 340 K. The integrated intensity of the (002) reflection in the X phase (I$_{\rm{X}}$) is 1.5 times larger than that in the T-phase (I$_{\rm{T}}$), and five times larger than that in the C-phase (I$_{\rm{C}}$) (${\rm I}_{\rm{X}}:{\rm I}_{\rm{T}}:{\rm I}_{\rm{C}} \sim 5:3.3:1$), which indicates that the extinction effect~\cite{Extinction} is phased out through transitions. In the zero-field heating process (ZFH), the X $\rightarrow$ T phase transition takes place at 380 K with a large hysteresis; the transition temperature is 40 K higher than that of the cooling process, while the T $\rightarrow$ C phase transition shows a small hysteresis $\sim$10 K. 

\subsection{Temperature dependence of lattice constants in FC process}

%%%%
%%%% ============================= Fig. 5 ============================== %
%%%%
\begin{figure*}[t]
\begin{center}
\includegraphics[width=0.8\linewidth]{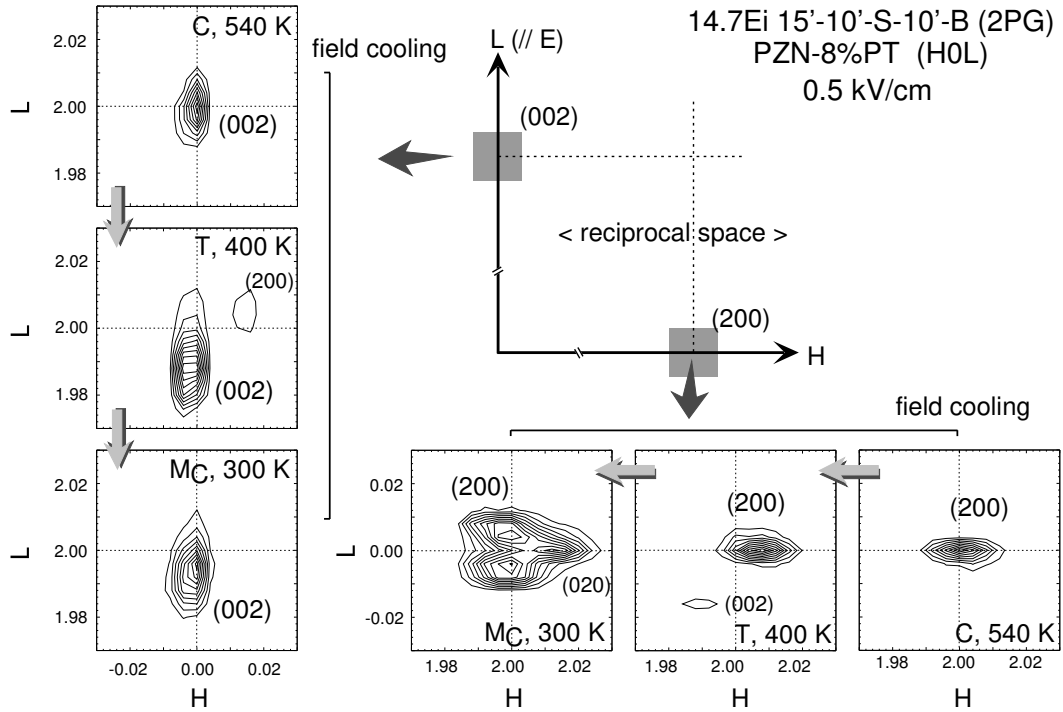}
\end{center}
\caption{
Mesh scans around the (002) and (200) at 540 K under the zero-electric field $E$ = 0.0 kV/cm and at 400 K and 300 K under $E$ = 0.5 kV/cm.
}
\label{fig:5}
\end{figure*}
%%%% ============================= Fig. 5 ============================== %

%%%%
%%%% ============================= Fig. 6 ============================== %
%%%%
\begin{figure}[t]
\begin{center}
\includegraphics[width=0.9\linewidth]{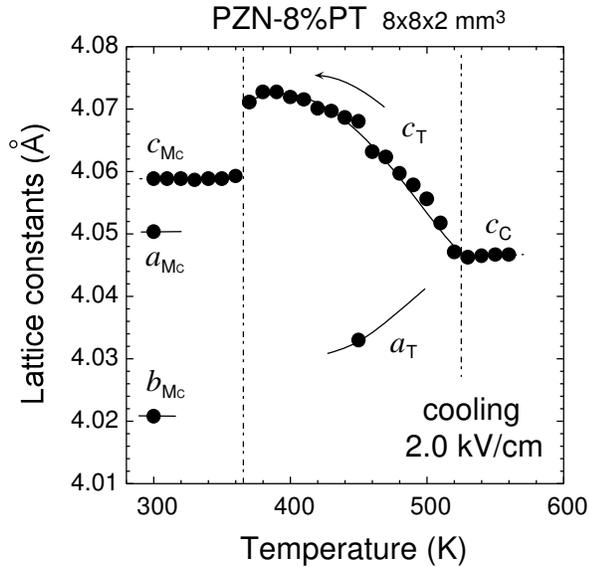}
\end{center}
\caption{
Temperature dependence of the lattice constants under the electric field $E$ = 2.0 kV/cm. The M$_{\rm{C}}$ phase is stable at RT in contrast to the ZF cooling result. Solid lines drawn through the data points are for guides to the eyes.
}
\label{fig:6}
\end{figure}
%%%% ============================= Fig. 6 ============================== %

  To clarify the electric field effects on the phase transition sequence as represented in Fig.~\ref{fig:2}(a), we studied the temperature dependence of the lattice constants of PZN-8\%PT crystal under fields ($E$ = 0.5, 2.0, 6.0, 10.0 kV/cm) on cooling, thus FC processes.
  To obtain a comprehensive picture of the PZN-8\%PT structural properties under the FC process, we took mesh scans around the (002) and (200) reflections at 540 K, 400 K and 300 K with the electric field $E$ = 0.5 kV/cm (see Fig.~\ref{fig:5}). As mentioned in Section II, we fixed the $a^{*}$ and $b^{*}$ values at 1.555 \AA$^{-1}$ throughout the scans. The followings are remarks of the mesh scan results. (a) At 540 K, neither elongation nor contraction of the lattice constants was seen. The PZN-8\%PT system is still in the cubic symmetry. (b) At 400 K, the elongation of the lattice constant $c_{\rm{T}}$ and the contraction of the lattice constant $a_{\rm{T}}$ were clearly seen. Thus the PZN-8\%PT crystal transforms into the T phase. 
%%%%%%%%%%
As shown in the contour maps of Fig.~\ref{fig:5}, additional peaks appears near (002) and (200) in the T phase. These peaks indicate a 90$^{\circ}$-domain formation along the \{101\} pseudo-cubic plane~\cite{Jona}.
%%%%%%%%%%
  (c) The M$_{\rm{C}}$ phase appears. As shown in Fig.~\ref{fig:5}(c), the (200) peak split into three peaks i.e.\ (200) twin peaks and one (020) single peak, while the (002) peak remains as a single peak.

  Figure~\ref{fig:6} shows the temperature dependence of the lattice constants observed at $E$ = 2.0 kV/cm. We also measured the (200) reflection at 450 K, the (200) and the (020) reflections at 300 K to obtain the information of the lattice constant $a_{\rm{T}}$ at 450 K, $a_{\rm{M_{\rm{C}}}}$ and $b_{\rm{M_{\rm{C}}}}$ at 300 K respectively. The lattice constant $c_{\rm{T}}$ gradually increases as the temperature decreases and suddenly drops largely at 340 K, where the T $\rightarrow$ M$_{\rm{C}}$ phase transition takes place.

\subsection{Electric field dependence of lattice constants at fixed temperatures}

%%%%
%%%% ============================= Fig. 7 ============================== %
%%%%
\begin{figure}[t]
\begin{center}
\includegraphics[width=0.85\linewidth]{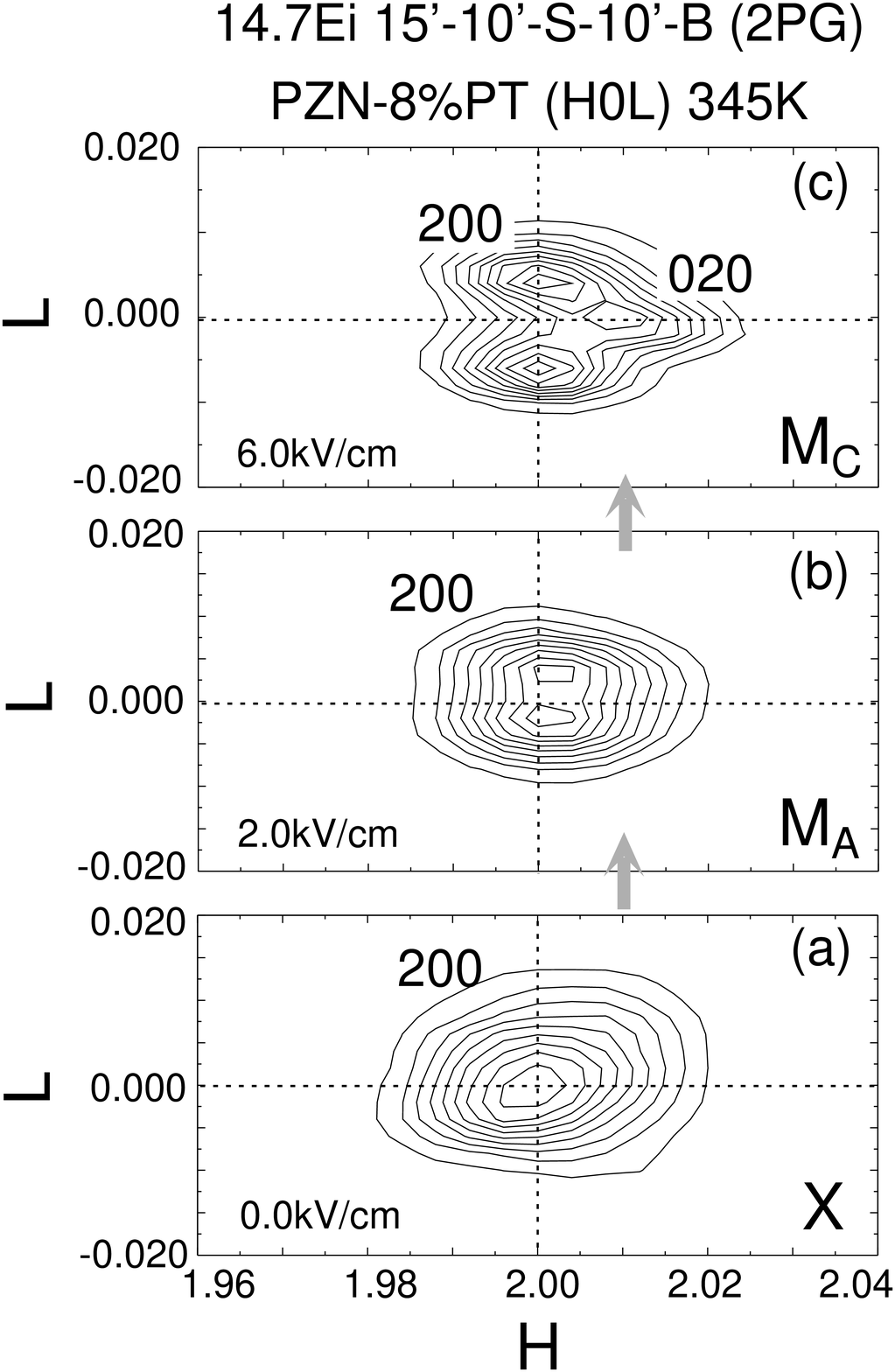}
\end{center}
\caption{
Electric field dependence of the (H0L) contour around the pseudo-cubic (200) obtained at 345 K.
}
\label{fig:7}
\end{figure}
%%%% ============================= Fig. 7 ============================== %

%%%%
%%%% ============================= Fig. 8 ============================== %
%%%%
\begin{figure}[t]
\begin{center}
\includegraphics[width=0.9\linewidth]{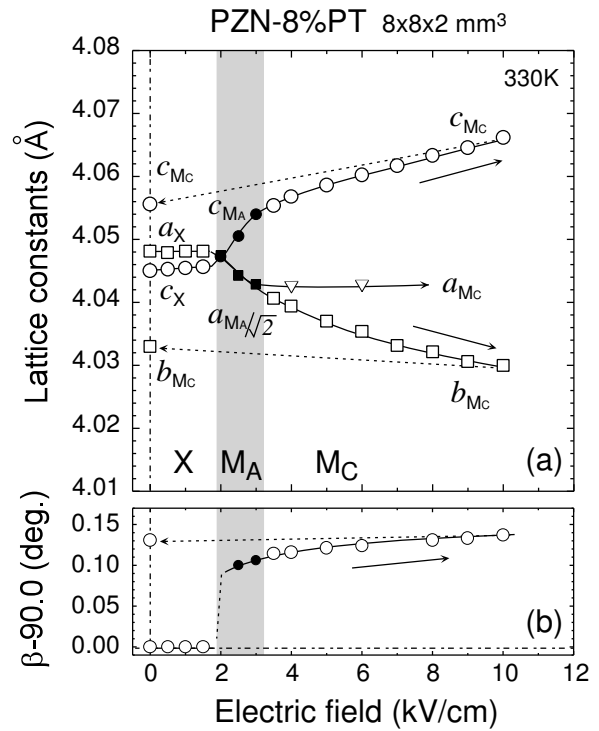}
\end{center}
\caption{
Electric field dependence of (a) the lattice constants and (b) $\beta-90^{\circ}$ observed at 330 K. Solid lines drawn through the data points are for guides to the eyes.
}
\label{fig:8}
\end{figure}
\begin{figure}[t]
%%%% ============================= Fig. 8 ============================== %

%%%%
%%%% ============================= Fig. 9 ============================== %
%%%%
\begin{center}
\includegraphics[width=0.9\linewidth]{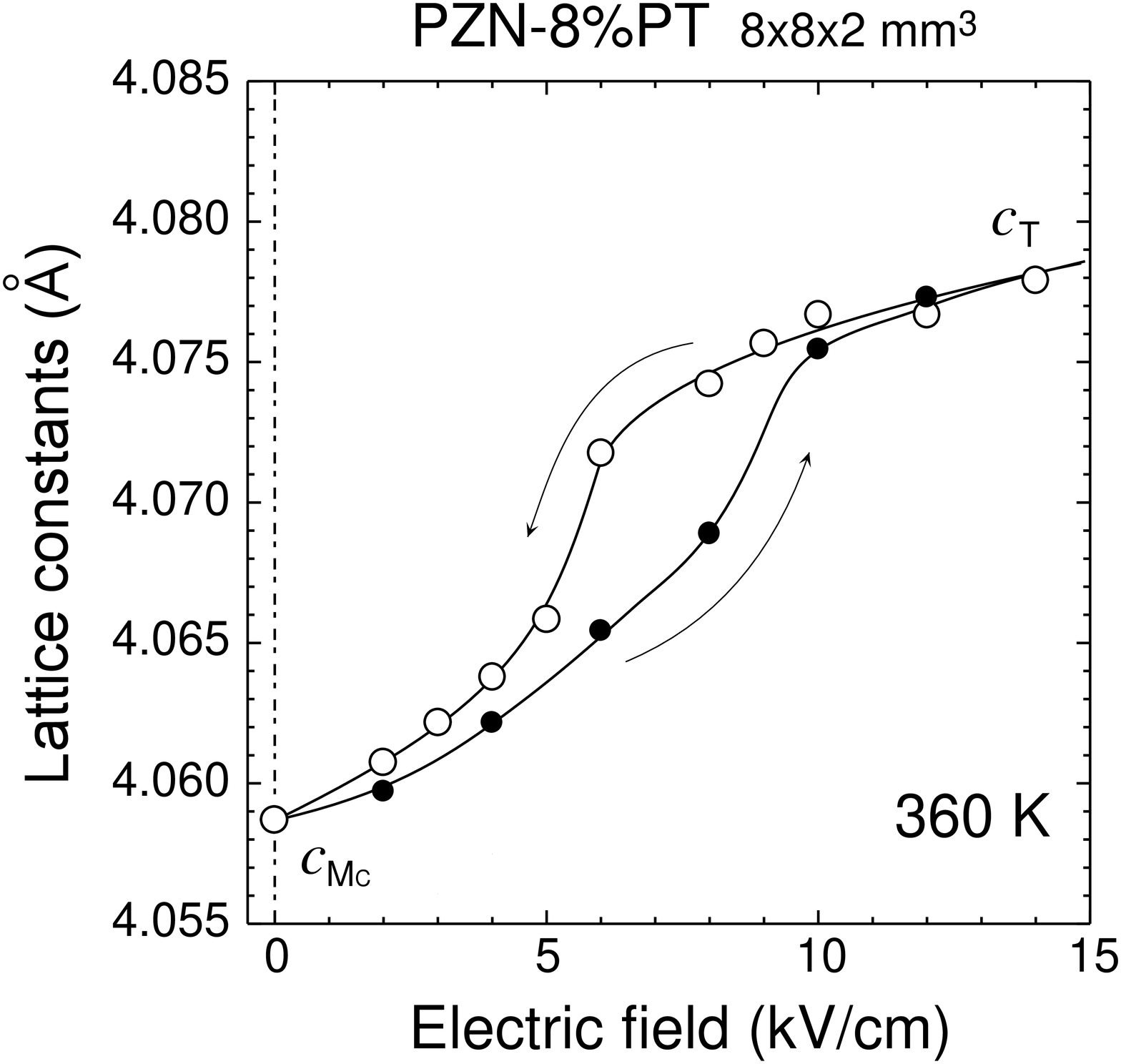}
\end{center}
\caption{
A hysteresis loop of the lattice constant $c$ representing the M$_{\rm{C}}$ $\rightarrow$ T phase transition of PZN-8\%PT. Solid lines drawn through the data points are for a guide to the eyes.
}
\label{fig:9}
\end{figure}
%%%% ============================= Fig. 9 ============================== %

  To construct Fig.~\ref{fig:2}(b), we also observed the electric field ($E$) dependence of the lattice constants at some selected temperatures: (1) From the C phase at 550 K and 520 K, (2) From the X phase at 300 K, 330 K, 345 K, and 360 K. The X phase in (2) was realized by FC from 550 K to 300 K followed by FH to the target temperature. Note in particular that PZN-8\%PT is in the T $\rightarrow$ X hysteresis loop at 360 K. Following the precedent routine, we also took a (H0L) mesh scan around the pseudo-cubic (200) position. As mentioned before, the peak structure around (200) strongly depends on the nature of the phase appearing in PZN-8\%PT. Figure~\ref{fig:7} shows the electric field dependence obtained at a fixed temperature of 345 K.
  We observed the irreversible X $\rightarrow$ M$_{\rm{A}}$ $\rightarrow$ M$_{\rm{C}}$ transition sequence which was found out by Noheda {\it et al.}~\cite{Noheda1,Noheda2,Ohwada}. The M$_{\rm{A}}$ phase is stable at room temperature for more than two weeks after the field removal, and shows no relaxation of the crystal lattice.
  
  Figure~\ref{fig:8} shows the electric field dependence of (a) the lattice constants and (b) the $\beta-90^{\circ}$ value, observed at 330 K. $a_{\rm{X}}$ and $c_{\rm{X}}$ show no field dependence. Above $E$ = 2.0 kV/cm, on the other hand, PZN-8\%PT transforms into the M$_{\rm{A}}$ phase with a large elongation of $c_{\rm{M_{\rm{A}}}}$ and shrink of $a_{\rm{M_{\rm{A}}}}$~\cite{Remark1}. 
  A peak split suddenly takes place at 2.0 kV/cm and $\beta-90^{\circ}$~\cite{Remark2} shows a gradual increment as the field increases. The M$_{\rm{A}}$ phase exists only in the range 2.0 $\le$ $E$ \textless 3.5 kV/cm, which is shown by gray region in Fig.~\ref{fig:8}. The width of the omega scan~\cite{Remark3} $\Delta\omega$, which corresponds to the crystal mosaic width, is also drastically changed from $\Delta\omega \sim0.2^{\circ}$ to $\sim0.1^{\circ}$ within the M$_{\rm{A}}$ phase as the field increases. The field dependent behavior of the lattice constants $c_{\rm{M_{\rm{A}}}}$ and the $a_{\rm{M_{\rm{A}}}}$ observed in the field range 2.0 $\le$ $E$ \textless 3.5 kV/cm seemed to be non-linear to the electric field, which is usually seen in the domain rotation properties by the bulk measurements~\cite{LP}. Since we observed the lattice constants, this anomalous property of the lattice constants dose not originate from the macroscopic domain rotation.
  
  Above 3.5 kV/cm, the M$_{\rm{A}}$ phase transforms into the M$_{\rm{C}}$ phase with a gradual change in spite of the polarization jump from the intra pseudo-cubic \{1$\overline{1}$0\} plane to intra pseudo-cubic \{010\} plane (see Fig.~\ref{fig:1}(b)). Up to 10 kV/cm, the lattice constant $c_{\rm{M_{\rm{C}}}}$ elongates and the $b_{\rm{M_{\rm{C}}}}$ shrinks continuously. On the other hand, the lattice constant $a_{\rm{M_{\rm{C}}}}$ does not show the field dependence. The $\beta-90^{\circ}$ value also represents a gradual increment up to 10 kV/cm. After removing the field, the lattice constants do not return to the home position, but still remain in the M$_{\rm{C}}$ symmetry. It is now clear that the M$_{\rm{C}}$ phase is another ground state of PZN-8\%PT at 330 K.

  Next we measured the $c$-axis jump ($c$-jump) on the poled PZN-8\%PT crystal at 360 K which is located inside the hysteresis loop of the reversible T $\rightarrow$ M$_{\rm{C}}$ phase transition. A sharp $c$-jump was clearly seen by strain and x-ray measurements at room temperature at high electric field $\sim$15 kV/cm as shown in Fig.~\ref{fig:1}(a). As mentioned before, Noheda {\it et al.}\ clarified using high energy x-rays that the crystal surface behaves differently from the crystal bulk at RT~\cite{Noheda2}, i.e.\, the field inducing the M$_{\rm{C}}$ $\rightarrow$ T transition is 7 kV/cm for the bulk and 20 kV/cm for the near surface region~\cite{Noheda2}. Neutrons easily penetrate a crystal and give us the bulk information more than any other measurements. Figure~\ref{fig:9} shows the hysteresis loop of the $c$ lattice constant, which represents the field induced M$_{\rm{C}}$ $\rightarrow$ T phase transition of PZN-8\%PT. The $c_{\rm{M_{\rm{C}}}}$ gradually increases non-linearly and shows a small jump, which is contrary to the one as seen in Fig.~\ref{fig:1}(a). This is ascribed to a strain distribution in the bulk~\cite{Ohwada}.

%%%
%%% Discussions
%%%

\section{Discussions}

\subsection{X phase}

  Let us first discuss the X phase introduced in Section III. 
  As shown in Fig.~\ref{fig:8}(a), the lattice constants $a_{\rm{X}}$ and $c_{\rm{X}}$ at $E$ = 0.0 kV/cm show a significant difference as much as $\Delta \sim$ 0.07\%. This difference is also seen in the (200) and (002) peak profiles of the C phase and X phase as depicted in Fig.~\ref{fig:4}. This means that the PZN-8\%PT crystal is {\it not} in the Rhombohedral symmetry after the ZFC process as it has been long believed. An important question is how PZN-8\%PT can distinguish between the $a_{\rm{X}}$ and $c_{\rm{X}}$ directions under the ZFC process.
%%%
This is most likely due to a sample treatment effect such as poling, quenching, attachment of electrodes to the $c$ plane, sample holding and so on. It would be interesting to study the effects on the X phase formation.
    
  As for the microscopic mechanism, a phase shift model proposed by Hirota {\it et al.}~\cite{Hirota} for the relaxor PbMg$_{1/3}$Nb$_{2/3}$O$_{3}$ (PMN) would give a clue to understand the X phase. They analyzed the diffuse intensity of PMN and found that the ionic displacements are divided into two categories; a phase shift ($\delta_{shift}$) common to all the ions and each ionic displacement which satisfies the center of mass condition as is the case for ordinary ferroelectric systems such as PbTiO$_{3}$. It is speculated that the inhomogeneity of Mg$^{2+}$ and Nb$^{5+}$ distribution creates a local electric field gradient and interacts with Polar Nano Regions resulting in such a phase shift. If the Mg$^{2+}$ and Nb$^{5+}$ distribution orients to a certain direction by a sample treatment as presented above, the phase shift is also orients and results in the anisotropy of the crystal lattice. We speculate that this type of phase shift also exists in the PZN-8\%PT crystal and causes a significant difference between the $a_{\rm{X}}$ and $c_{\rm{X}}$ lattice constants. 

%%%%%%%
%As represented in Section III, the 90$^{\circ}$-domains are formed in the T phase with the domain walls running along the \{100\} pseudo-cubic plane (90$^{\circ}$-domain$^{\{100\}}$) in the ZFC process. Once suppressed, the X phase does not appear on cooling even after the field is removed. We thus speculated that the strain field from the 90$^{\circ}$-domain$^{\{100\}}$ boundaries are closely related to the formation of the X phase.
%%%%%%%

  One of the main goal for the relaxor physics is to understand and calculate how a large piezoelectric response is realized. The discovery of the novel X phase as a ground state of PZN-8\%PT around RT urges revisions of theoretical frameworks of relaxors. In particular, the polarization rotation mechanism\cite{Fu}, which successfully explains the high electromechanical responses, is largely based on the correct assignment of the symmetries of various structural phases. It is thus very important to identify the space group of the X phase.

\subsection{$E$-$T$ phase diagram}

  As described in Section III, we have constructed the $E$-$T$ phase diagram of PZN-8\%PT. Figure~\ref{fig:2}(a) summarizes the results obtained in the FC process. The C $\rightarrow$ T phase boundary shifts to higher temperature and the T $\rightarrow$ M$_{\rm{C}}$ phase boundary shifts slightly to lower temperature as the field increases, which indicates that the T phase is stabilized under a high electric field. The M$_{\rm{A}}$ phase never appear in the FC process even when the field is very low, as experimentally confirmed down to $E$ = 0.5 (kV/cm). Although it is not shown in Fig.~\ref{fig:2}(a), the M$_{\rm{C}}$ phase is stable even at 15 K. Figure~\ref{fig:2}(b) summarizes the field increasing process at a fixed temperature after ZFC (and ZFH for X phase). At high temperature, the C phase transforms into the T phase reversibly and the phase boundary shows the same feature as that in Fig.~\ref{fig:2}(a). At low temperature, the X phase irreversibly transforms into the M$_{\rm{A}}$ phase at a lower field. The M$_{\rm{A}}$ phase is represented by the gray region in Fig.~\ref{fig:2}(b). Note that the X and M$_{\rm{A}}$ phases become stable as the temperature decreases. As already stated above, the M$_{\rm{A}}$ phase does not relax at least for two weeks. The PZN-8\%PT crystal further transforms into the M$_{\rm{C}}$ phase at a higher field irreversibly. Once the M$_{\rm{C}}$ phase is obtained (PZN-8\%PT poled crystal), the X and M$_{\rm{A}}$ phases no longer recover. Only the reversible M$_{\rm{C}}$ $\leftrightarrow$ T phase transition takes place at higher field as seen in the strain field (Fig.~\ref{fig:1}(a)) and x-ray measurements
  
  It has been reported that the PZN-9\%PT crystal has the orthorhombic (O) symmetry at RT after applying field~\cite{Uesu}. Let us consider a new phase diagram which has the composition ($x$) axis perpendicular to the $E$-$T$ plane. In that phase diagram, it is clear that the O phase is indeed very close to the M$_{\rm{C}}$ phase. We speculate that a {\it hidden} O symmetry exists in the PZN-8\%PT at RT and gives birth to the M$_{\rm{C}}$ symmetry, which polarization is located on the T-M$_{\rm{C}}$-O polarization rotation path as shown in Fig.~\ref{fig:1}(b). Park {\it et al.} reported that, in the PZN-$x$PT system, the M$_{\rm{C}}$ symmetry near MPB gives a higher piezoelectricity than the M$_{\rm{A}}$ symmetry (lower $x$) does~\cite{Park1}. This difference can be ascribed to the background symmetry of each phase: the O symmetry for M$_{\rm{C}}$ and the R symmetry for M$_{\rm{A}}$. The O symmetry requires the [001] electric field much higher than that for the R symmetry, resulting in the higher piezoelectricity of the M$_{\rm{C}}$ phase. 

%%%
%%% Summary
%%%

\section{Summary}

  We made comprehensive neutron-diffraction studies on high-temperature and high-electric-field  effects on Pb[(Zn$_{1/3}$Nb$_{2/3}$)$_{0.92}$Ti$_{0.08}$]O$_{3}$ (PZN-8\%PT) in the ranges $300 \le T \le 550$~K and $0 \le E \le 15$~kV/cm. In the field cooling process (FC, $E \ge 0.5$~kV/cm), a successive cubic (C) $\rightarrow$ tetragonal (T) $\rightarrow$ monoclinic (M$_{\rm C}$) transition was observed. In the zero field cooling process (ZFC), however, we have found that the system does {\em not} transform to the rhombohedral (R) phase as widely believed, but to a new, unidentified phase, which we call X. X gives a Bragg peak profile similar to that expected for R, but the $c$-axis is always slightly shorter than the $a$-axis. We expect that the discovery of the novel X phase as a ground state of PZN-8\%PT urges revisions of theoretical frameworks of relaxors. As for field effects on the X phase, we found an irreversible X $\rightarrow$ M$_{\rm C}$ transition via another monoclinic phase (M$_{\rm A}$) as expected from a previous report~\cite{Noheda1}. At a higher electric field, we confirmed a $c$-axis jump associated with the field-induced M$_{\rm C}$ $\rightarrow$ T transition, as observed by strain and x-ray diffraction measurements. Our precise $E$-$T$ phase diagram will provide a fundamental aspect for future studies of relaxors. Our next goal is to clarify a true character of the X phase.

  After this work was completed, we have learned that a novel phase, which is similar to the X phase of PZN-8\%PT, was found in PZN at room temperature by high energy x-ray diffraction~\cite{xu}.

%%%
%%% Acknowledgments
%%%

\begin{acknowledgments}

  We would like to thank P.\ M.\ Gehring, S-E.\ Park, B.\ Noheda and S.\ Wakimoto for stimulating discussions, as well as Y. Kawamura for technical support. We are also grateful to G. Xu and his collaborators for informing us of their high energy x-ray results on PZN prior to publication. This work was supported by U.S.-Japan Cooperative Research Program on Neutron Scattering between the U.S.\ Department of Energy (U.S.DOE) and the Japanese MONBU-KAGAKUSHO. We also acknowledge financial support from the U.S.DOE under Construction No. DE-AC02-98CH10886.

\end{acknowledgments}


\begin{thebibliography}{99}
\bibitem{M} The polarization vectors lie in the pseudo-cubic \{1$\overline{1}$0\} (\{100\}, \{010\}) plane in M$_{\rm{A}}$ (M$_{\rm{B}}$, M$_{\rm{C}}$). This notation is proposed by Vanderbilt and Cohen~\cite{VC}.
\bibitem{Remark1} The crystal lattice of the M$_{\rm{A}}$ phase is defined as
$\vec{a_{\rm{M_{\rm{A}}}}} = \vec{a_{\rm{C}}}+\vec{b_{\rm{C}}}$,
$\vec{b_{\rm{M_{\rm{A}}}}} = \vec{a_{\rm{C}}}-\vec{b_{\rm{C}}}$ and
$\vec{c_{\rm{M_{\rm{A}}}}} = \vec{c_{\rm{C}}}$~\protect{\cite{Noheda2}}, where $\vec{a_{\rm{C}}}$, $\vec{b_{\rm{C}}}$ and $\vec{c_{\rm{C}}}$ are unit vectors of a cubic lattice.
\bibitem{Remark2} In Fig.~\ref{fig:7}(b), the $\beta$ angle of each monoclinic phase is related to the real monoclinic angle $\beta_{M_{\rm{A}}}$ and $\beta_{M_{\rm{C}}}$ as $\tan\beta \sim \sqrt2\tan\beta_{M_{\rm{A}}}$ and $\beta = \beta_{M_{\rm{C}}}$.
\bibitem{Remark3} Omega scan is one which goes perpendicularly to the in-plane scattering vector.
\end{thebibliography}
\end{document}